\documentclass[aps,twocolumn,showpacs,eqsecnum]{revtex4}

\usepackage{color}
\usepackage{fancybox}
\usepackage{epsfig}

\usepackage{latexsym}
\usepackage{amsmath}
\usepackage{xspace}
\usepackage{amssymb}

\newcommand{\bra}[1]{\langle #1|}
\newcommand{\ket}[1]{| #1 \rangle }
\newcommand{\Bra}{\langle}

\newcommand{\Swap}[1]{{\rm Swap_{#1}}}

\newcommand{\as}[1]{{\rm ASym}_{#1}}
\newcommand{\C}{{\rm \Pi}}
\newcommand{\E}{{\rm E}}
\newcommand{\Hs}{{\cal H}}
\newcommand{\one}{{\openone}}
\newcommand{\nstav}{{\rm \alpha_{?}}}
\newcommand{\pui}[3]{P_{#1}(\ket{#2},\ket{#3})}
\newcommand{\mpui}[2]{\overline{P_{#1}(#2)}}

\newcommand{\complex}{{\mathbb C}}

\begin{document}
\title{Unambiguous coherent state identification: Searching a quantum database}
\author{Michal Sedl\'ak$^{1}$,
M\'ario Ziman$^{1,2}$,
Ond\v rej P\v ribyla$^{3}$,
Vladim\'\i r Bu\v zek$^{1,4}$,
and Mark Hillery$^{5}$}
\title{Unambiguous identification of coherent states: Searching a quantum database}
\address{
$^{1}$Research Center for Quantum Information, Slovak Academy of Sciences,
D\'ubravsk\'a cesta 9, 845 11 Bratislava, Slovakia \\
$^{2}$Faculty of Informatics, Masaryk University, Botanick\'a 68a,
602 00 Brno, Czech Republic\\
$^{3}$Faculty of Science, Masaryk University, Kotl\'{a}\v{r}sk\'{a} 2,
611 37 Brno, Czech Republic\\
$^{4}${\em Quniverse}, L{\'\i}\v{s}\v{c}ie \'{u}dolie 116, 841 04
Bratislava, Slovakia\\
$^{5}$Department of Physics and Astronomy, Hunter College
of the City University of New York, 695 Park Avenue, New York, NY 10021, USA
}
\begin{abstract}
We consider an unambiguous identification of an {\em unknown}
coherent state with one of two {\em unknown} coherent reference
states. Specifically, we consider two modes of an electromagnetic field
prepared in {\em unknown} coherent states $|\alpha_1\rangle$ and $|\alpha_2\rangle$, respectively.
The third mode is prepared either in the state $|\alpha_1\rangle$ or in the state $|\alpha_2\rangle$.
The task is to identify (unambiguously) which of the two modes are in the same state.
We present a scheme consisting of three beamsplitters capable to perform
this task. Although we don't prove the optimality, we show that the
performance of the proposed setup is better than the
generalization of the optimal measurement known for a finite-dimensional case.
We show that a single beamsplitter is capable
to perform an unambiguous quantum state comparison for coherent states
optimally. Finally we propose an experimental setup consisting of
$2N-1$ beamsplitters for unambiguous identification among
$N$ {\em unknown} coherent states. This setup can be considered as a search in a quantum database. The elements of the
database are {\em unknown} coherent states encoded in different modes of an electromagnetic field. The task is to specify
the two modes that are excited in the same, though unknown, coherent state.

\end{abstract}

\maketitle

\section{Introduction}
The ability to discriminate quantum states plays an important
role in  quantum information processing. Because of the
quantum interference two (non-orthogonal) quantum states cannot be
distinguished perfectly providing the number of copies of these
states is limited. The topic of quantum state discrimination
was firmly established in 1970s by pioneering work of Helstrom
\cite{helstrom}, who considered a minimum error discrimination
of two known quantum states. In this case the state identification
is probabilistic. Another equally significant
approach is the unambiguous discrimination of quantum states,
originally formulated and analyzed by Ivanovic, Dieks and Peres
\cite{ivanovic,dieks,peres} in 1987. In contrast to the minimum error
discrimination approach,  the unambiguous state identification
is deterministic, i.e. no erroneous conclusions
are permitted. But in addition an inconclusive result is allowed
corresponding to situations in which the state identification fails.
The solution for unambiguous discrimination of two {\em known} pure states
appearing with arbitrary prior probabilities (further denoted
as $\eta_1, \eta_2$) was obtained by Jaeger and Shimony \cite{jaeger}.
Subsequent research was mainly focused on unambiguous discrimination
among several known pure states and unambiguous discrimination
of two mixed states, which is still an open problem.
The physical implementation of the optimal unambiguous discriminator
device working for arbitrary coherent states was proposed
by K. Banazsek \cite{banaszek}.

All results mentioned above are heavily based on a
prior classical knowledge we have about those quantum states that are to
be discriminated. S. Barnett {\it et al.} \cite{jex}
have studied this problem in detail and in addition they have addressed
the following intriguing question: Is it possible
to say anything unambiguously whether pure quantum
states of a pair of identical quantum systems (finite
dimensional) are equal or not? Here no prior knowledge about
the states is assumed.
This problem is called quantum state comparison
(for extension to more systems see Ref.~\cite{jex04}). It turned out that
symmetry with respect to the exchange of the subsystems enables
one to reveal unambiguously the difference between the states of the
subsystems, however their equality cannot be determined
unambiguously. E. Andersson {\it et al.} \cite{andersson}
investigated this problem also for coherent states and proposed
a simple setup (described in Sec.~II), consisting of one beam splitter
and a photodetector, capable to perform this task.

Such results stimulated J. Bergou and M. Hillery \cite{bergou} to
reduce the prior knowledge in unambiguous discrimination of quantum
states. These authors formulated the following problem:
Imagine we are given two qubits $B$ and $C$ each of them in an
unknown pure state. At the same time we are given also a third
qubit $A$, which is guaranteed to be either in the state of the
first or the second qubit. The task is to determine unambiguously
with which of the two qubits the state of the third qubit matches.
In such modification of the original problem the whole
information is conveyed by states of quantum systems.
Given the fact that we have just a single copy of each of the two states
even the optimal quantum mechanical measurement would allow us
to determine those states with a rather small fidelity (for more details
see Refs.~\cite{helstrom,massar,derka}). Therefore, instead of trying to estimate
reference states we directly use them in their ``quantum'' form.
The control of the behavior of the device by the
quantum program register is the key feature of programmable quantum
devices \cite{nielsen,hillery01,dusek} (for a review see Ref.~\cite{buzek2006}).
In these devices one subsystem serves as the data register and the
second quantum subsystem as the program register that carries instructions
about transformations the machine has to perform on the data register.

The problem of unambiguous discrimination of unknown states, called
also the unambiguous quantum state identification problem (UI), has been
investigated over last few years by many authors.
Bergou {\it et al.} \cite{bergou2} examined the situation
with more copies of the third qubit $A$.
If qudits ($d$ dimensional quantum systems) instead of the qubits
are used and more copies of qudits A and B are provided
then the analytical solution for equal prior probabilities was
obtained by A. Hayashi {\it et al.} in Ref.~\cite{hayashi}.
C. Zhang and M. Ying \cite{zhang} investigated the unambiguous
identification among $n$ unknown qudit states and they provided
necessary and sufficient criterion characterizing all possible
programmable discriminators performing this task.

One of the aim of this paper is to illustrate that a prior knowledge of
a subset of states ${\cal S}$ uniformly entering subsystems $A$ and
$B$ can significantly affect the optimal UI measurement and the performance
of it with respect to the universal UI measurement. We illustrate this
on two examples: The first example deals with the so called  equatorial qubits described
in Sec. II.B. The second example deals with coherent states
examined in Sec. IV. In Sec. III we present how an ``intuitive'' universal UI measurement
of systems of an arbitrary dimension can be constructed. Further we will compare
different types of UI devices with the optimal
universal one found in Ref.~\cite{hayashi}. Although the UI measurement
of coherent states will not be proved to be optimal,
 we will show that it rapidly outperforms the universal
UI measurement. In addition we will show that in the case of continuous variable when the inputs are
represented by coherent states the UI measurement  can be also easily implemented
by three beam splitters and two photo-detectors.

Formally the unambiguous identification problem (UI)
fits into the following framework: Three identical
subsystems A,B and C are prepared in unknown pure product states
$\ket{\psi_?}_A$, and $ \ket{\psi_1}_B,\ket{\psi_2}_C$, respectively.
Furthermore, the subsystem A is guaranteed to be either in the same state
as the subsystem B or as the subsystem C.
Thus two types of states should be discriminated:
\begin{eqnarray}
\nonumber
\ket{\Psi_1}_{ABC}\equiv\ket{\psi_1}_{A}\otimes\ket{\psi_1}_{B}\otimes\ket{\psi_2}_{C}\, ,\\
\ket{\Psi_2}_{ABC}\equiv\ket{\psi_2}_{A}\otimes\ket{\psi_1}_{B}\otimes\ket{\psi_2}_{C}\, .
\label{stavy1}
\end{eqnarray}
The unambiguous identification machine is described
by means of a positive operator value measure (POVM)
consisting of three elements $\{E_0,E_1,E_2\}$.
Element $\E_1$ (respectively $\E_2$) corresponds to correct
identification of $\ket{\Psi_1}$ (respectively $\ket{\Psi_2}$)
type of state and $\E_0$ corresponds to the inconclusive result.
These elements must obey no-error conditions [Eq.(\ref{stavy2})]
and constitute a proper POVM [Eq.(\ref{povm})]:
\begin{eqnarray}
Tr[\E_1\rho_2]=Tr[\E_2\rho_1]=0\, ; \quad
\rho_i =\ket{\Psi_i}\bra{\Psi_i} \, ;\label{stavy2}\\
\E_i\geq0,\E_0\geq0\, ; \quad
\E_0+\E_1+\E_2=\one\, .
\label{povm}
\end{eqnarray}
It is assumed that the state
$\ket{\Psi_1}$ ($\ket{\Psi_2}$) appears with a prior
probability $\eta_1$ ($\eta_2$). The performance of the UI measurement
is quantified by a probability of identification for a particular
choice of states
\begin{equation}
\pui{}{\psi_1}{\psi_2}=\eta_1 Tr[\E_1\rho_1]+\eta_2 Tr[\E_2\rho_2] \label{pidentify}\, .
\end{equation}
Although $\pui{}{\psi_1}{\psi_2}$ is not a measurable quantity
in the problems we consider, it will be very useful for comparison
of different UI measurements. Alternatively, we can use
the average value to evaluate the performance of UI devices
\begin{eqnarray}
\mpui{}{S}=\int_S \int_S \pui{}{\psi_1}{\psi_2} \quad d\psi_1 d\psi_2 \, .
\label{meanpi}
\end{eqnarray}
In what follows we will denote the set of all pure states
of a $d$-dimensional quantum system (qudit) by $S_d$ and the
subscript of $P$ will indicate the used UI measurement.
The optimality of UI measurements is defined with respect to
their average performance, i.e. the aim is to optimize $\mpui{}{S}$.

\section{Unambiguous Identification for qubits}
The solution to the unambiguous identification problem for a qubit was given
in Ref.~\cite{bergou}. The optimal UI measurement depends only on prior
probabilities $\eta_1=\eta$, $\eta_2=1-\eta$. Specifically, there are three different
regions of values of $\eta$ for which the POVM operators are specified as follows:
\begin{equation}
\begin{array}{cll}
0\le\eta<\frac{1}{5}: & \E^{opt}_1=0\, ; & \E^{opt}_2=\one_C\otimes\psi^-_{AB}\, ; \\
\frac{1}{5}\le\eta\le\frac{4}{5}\, : &
\E^{opt}_1=\lambda\one_B\otimes\psi^-_{AC}\, ; &
\E^{opt}_2=\frac{4-4\lambda}{4-3\lambda}\one_C\otimes\psi^-_{AB}\, ;
\\
\frac{4}{5}<\eta\le 1\, :
& \E^{opt}_1=\one_B\otimes\psi^-_{AC}\, ; \quad
& \E^{opt}_2=0\, ,
\end{array}
\nonumber
\end{equation}
where $\ket{\psi^-_{AB}}=1/\sqrt{2}(\ket{01}-\ket{10})_{AB}$,
$\lambda=\frac{2}{3}(2-\sqrt{\frac{\eta_2}{\eta_1}})$
and $\ket{\psi^-_{AB}}\bra{\psi^-_{AB}}=\psi_{AB}^-\equiv \as{AB}$
is a projector onto the antisymmetric part of the two qubit
Hilbert space $\Hs^{\otimes 2}_2$. The inconclusive result is associated
with the POVM element $\E^{opt}_0=\one-\E^{opt}_1-\E^{opt}_2$.

\subsection{Relation to quantum state comparison} \label{relcomparison}
Intuitively, the unambiguous state comparison and the unambiguous state
identification are very closely related problems. Indeed, one can consider
the following family of problems: given $n+1$ qudits in
unknown states $\psi_1,\dots,\psi_n,\psi_{n+1}$. All qudits [except the
$(n+1)$-th one] are guaranteed to be in different states. Decide whether
$(n+1)$-th qudit matches with one of the given qudits, or not.
The unambiguous state comparison \cite{jex} is a task with $n=1$
and unambiguous identification corresponds to $n=2$ providing that
the $(n+1)$-th qudit is promised to be in one of
the states $\psi_1,\dots,\psi_n$. Moreover, the
identification can be logically reduced to a series of
state comparisons, although such reduction does not necessarily
give an optimal identification scheme.
In particular, the UI measurement for $n=2$ can be always
designed out of the unambiguous comparison device in the following way:
The experimentalist randomly chooses (with probabilities $q$ and $1-q$, respectively)
one of the reference states ($|\psi_B\rangle,|\psi_C\rangle$)
to be the input of the unambiguous comparator machine together with
the unknown state $|\psi_A\rangle$. Performing the state comparison
$M_{AB}$ ($M_{AC}$) the conclusive result means that the unknown state
is $|\psi_C\rangle$ ($|\psi_B\rangle$)
Formally this corresponds to a probabilistic switching between two measurement
apparatuses resulting in the UI measurement consisting of POVM elements
$\overline{E}_1=q I_B\otimes F^{\rm dif}_{AC}$,
$\overline{E}_2=(1-q) I_C\otimes F^{\rm dif}_{AB}$,
$\overline{E}_0=I-\overline{E}_1-\overline{E}_2$, where
$F^{\rm dif}$ denotes the POVM element associated with the
conclusive result saying that two states are different \cite{jex}.

For a prior probability $\eta_1<\frac{1}{5}$
the optimal qubit UI measurement is projective,
distinguishing the antisymmetric and symmetric states of the subsystem $AB$.
Moreover, the qubit $C$ is not used at all, and the exchange symmetry of the
states with respect to systems $A$ and $B$ is measured distinguishing between
the states $\ket{\psi_1}_{A}\otimes\ket{\psi_1}_{B}$
and $\ket{\psi_2}_{A}\otimes\ket{\psi_1}_{B}$.
This is exactly the aim of a quantum state comparison of two unknown pure
states. States of the type $\ket{\psi}\otimes\ket{\psi}$ are from the symmetric
subspace, therefore the projection onto the antisymmetric subspace
unambiguously
identifies whether the states of the subsystems are different.
On the other hand projection onto the symmetric subspace is inconclusive,
because both types of states $\ket{\psi}\otimes\ket{\psi}$,
$\ket{\psi}\otimes\ket{\phi}$ have nonzero overlap with it.
Analogous considerations holds also for the interval
$\eta_1>\frac{4}{5}$ and a subsystem AC. For the aforementioned
prior probabilities the mean probability of identification
equals $\mpui{opt}{S_2}=\eta_{\min}/4$, where
$\eta_{\min}=\min\{\eta_1,\eta_2\}\in [0,1/5]$ and $S_2$ denotes the set
of all pure states of the qubit.

For equal prior probabilities the optimal measurement is a ``true'' POVM
measurement, whose elements $\E^{opt}_1$, $\E^{opt}_2$ are $2/3$ times the
above-mentioned quantum state comparison measurement
elements $\as{AC}$, $\as{AB}$. In this case the mean probability
of identification $\mpui{opt}{S_2}$ is $1/6$. Using the mixing strategy
the success probability can reach at most $1/8$.

\subsection{Unambiguous identification of equatorial qubits}
\label{eqqubits}
Let us consider a restricted set of pure states lying on the equator of the
Bloch sphere $\ket{\varphi}=1/\sqrt{2}(\ket{0}+e^{\imath\varphi}\ket{1})$
with $\varphi \in [0,2\pi)$. Let us denote the subset of all equatorial
states by $S_{eq}$. We are going to find the UI measurement, which optimizes
the probability of identification $\mpui{eq}{S_{eq}}$ averaged over the
set $S_{eq}$. Following the approach used in Ref.~\cite{bergou2}
we obtain
\begin{eqnarray}
\mpui{eq}{S_{eq}}=\eta_1 Tr[\E^{eq}_1\Omega_1]
+\eta_2 Tr[\E^{eq}_2\Omega_2]\, ,
\label{noverho}
\end{eqnarray}
with average states
\begin{eqnarray}
\Omega_j=\frac{1}{(2\pi)^2}
\int_{0}^{2\pi}\int_{0}^{2\pi}
\varphi_j\otimes\varphi_1\otimes\varphi_2\quad d\varphi_1 d\varphi_2\, ,
\end{eqnarray}
where we used the notation $\varphi_j=\ket{\varphi_j}\bra{\varphi_j}$.
After a little algebra this yields
\begin{equation}
\begin{aligned}
\Omega_1=\frac{1}{8}\one_C\otimes(\ket{00}\bra{00}+\ket{11}\bra{11}+2\ket{\psi^+}\bra{\psi^+})_{AB}\,
; \label{Omega1}
\\
\Omega_2=\frac{1}{8}\one_B\otimes(\ket{00}\bra{00}+\ket{11}\bra{11}+2\ket{\psi^+}\bra{\psi^+})_{AC}\,
,
\end{aligned}
\end{equation}
with $\ket{\psi^+}=\frac{1}{\sqrt{2}}(\ket{01}+\ket{10})$.
We integrate the no-error conditions (\ref{stavy2}) in the same way
and obtain $Tr[\E^{eq}_1\Omega_2]=Tr[\E^{eq}_2\Omega_1]=0$.
Operators $\E^{eq}_j$, $\Omega_j$ are positive therefore the previous
equation means that the operators in the trace have orthogonal supports.
Thus we determine the zero eigenvectors of the opeators $\Omega_1$ and $\Omega_2$,
$\Omega_2\ket{a_j}=0$ and  $\Omega_1\ket{b_j}=0$, respectively, where
\begin{eqnarray}
\ket{a_1}=\ket{0}_B\otimes\ket{\psi_{AC}^-}\, ; \quad \ket{b_1}=\ket{0}_C\otimes\ket{\psi_{AB}^-} \, ;\\
\ket{a_2}=\ket{1}_B\otimes\ket{\psi_{AC}^-}\, ; \quad \ket{b_2}=\ket{1}_C\otimes\ket{\psi_{AB}^-}\, .\nonumber
\end{eqnarray}
These eigenvectors
determine subspaces in which POVM elements $\E^{eq}_2$, $\E^{eq}_1$
\begin{eqnarray}
\E^{eq}_1=\sum^2_{j,k=1} \alpha_{jk} \ket{a_j}\bra{a_k}\, ; \quad \E^{eq}_2=\sum^2_{j,k=1} \beta_{jk} \ket{b_j}\bra{b_k} \label{e1equatorial}
\end{eqnarray}
can operate.
Our goal is to maximize $\mpui{eq}{S_{eq}}$, while keeping the POVM elements positive.
Therefore we use equations (\ref{Omega1}) and (\ref{e1equatorial}) to express equation (\ref{noverho}) only in
terms of the coefficients $\alpha_{jk}$, $\beta_{jk}$
\begin{eqnarray}
\mpui{eq}{S_{eq}}=\frac{\eta_1}{8}(\alpha_{11}+\alpha_{22})+\frac{\eta_2}{8}(\beta_{11}+\beta_{22}) \, .
\end{eqnarray}
Accidentally at the same time the expression for $\mpui{eq}{S_{eq}}$ coincides with $\mpui{opt}{S_2}$ and the states $\ket{a_j}$, $\ket{b_k}$
are the same as in Ref.~\cite{bergou2} [compare with Eqs. (3.19-3.22) of this reference].
Therefore the optimization task and the resulting
measurement is in our case exactly the same as for the universal UI of qubits. As a result we see that the optimal UI measurement for equatorial
states is the same as for the most general qubit state
(specified at the beginning of this section). Hence, in this case the a priori knowledge does not help us to improve the
performance of the UI measurement.

\section{Unambiguous Identification of Qudits}
\label{uiqudits}
\subsection{The Swap-based approach}
The POVM elements for the optimal universal UI of qubits
$\E^{opt}_{1},\E^{opt}_{2}$ are proportional to  the projectors
onto the antisymmetric subspace of the two-qubit subsystems AC and AB, respectively.
The simple generalization of the aforementioned universal UI measurement
to the case of qudits is the following POVM, which we abbreviate by
$sb$ (stands for the ``swap-based''):
\begin{equation}
\begin{aligned}
&\E^{sb}_{1}= c_{1} \one_{B}\otimes \as{AC}=c_{1}\frac{1}{2}(1-\Swap{AC})\, ;\\
&\E^{sb}_{2}= c_{2} \one_{C}\otimes \as{AB}=c_{2}\frac{1}{2}(1-\Swap{AB})\, ;\\
&\E^{sb}_{0}= \one-\E^{sb}_{1}-\E^{sb}_{2},
\label{sbpovm}
\end{aligned}
\end{equation}
where $\as{XY}$ denotes the projector to the
antisymmetric subspace of $X$ and $Y$ particle.

The positivity of $\E^{sb}_{0},\E^{sb}_{1},\E^{sb}_{2}$ results in the
conditions for $c_{1},c_{2}$. Namely, from
$\E^{sb}_1\ge 0,\E^{sb}_1\ge 0$ we obtain that $c_{1}\geq0$ and $c_{2}\geq 0$,
whereas the inequality imposed by the positivity of $\E^{sb}_{0}$
is not so apparent. Further we will calculate eigenvalues
of $\E^{sb}_{0}$ explicitly. Let $\{\ket{\,j\,}\}_{j=1}^{d}$ be
a basis of the qudit Hilbert space $\Hs_d$. Then
$\ket{ijk}\equiv\ket{i}_{A}\otimes\ket{j}_{B}\otimes\ket{k}_{C}$
is the basis of the three-qudit Hilbert space $\Hs^{\otimes3}_d$.
The operator $E^{sb}_{0}$ can be expressed in terms of the identity and the
$\Swap{}$ operators so $\bra{ijk}\E^{sb}_{0}\ket{lmn} \equiv 0$
whenever $\{ijk\}$ is not a permutation of $\{lmn\}$, i.e.
$ijk\ne \sigma(lmn)$. In other words if we properly reorder
the above basis the matrix $\E^{sb}_{0}$ is block diagonal.
The blocks are of the following three types:
\begin{itemize}
\item the trivial $1\times 1$ block $\bra{iii}\E^{sb}_{0}\ket{iii}=1$
\item the $3\times 3$ block with the matrix $\bra{\sigma_{1}(iij)}\E^{sb}_{0}\ket{\sigma_{2}(iij)}$
\item the $6\times 6$ block with the matrix $\bra{\sigma_{1}(ijk)}\E^{sb}_{0}\ket{\sigma_{2}(ijk)}$.
\end{itemize}
The dimensionalities of the blocks are given by the number of inequivalent permutations
$\sigma$ of the three indexes.

For qubits only the blocks of the first two types occur in the matrix $\E^{sb}_{0}$
whereas for qudits $(d>2)$ blocks of all three types arise.
Hence we reduced the problem of finding the eigenvalues
of the positive operator $\E^{sb}_{0}$ to an evaluation of the
eigenvalues of the matrices mentioned above, which is treated
in more details in Appendix A. It is shown there that the positivity of
$\E^{sb}_{0}$ imposes a particularly simple inequality
\begin{equation}
\label{so_simply!}
c_{1}+c_{2}\leq 1
\end{equation}
regardless of the dimension $d$ ($d>2$).

The probability $(\ref{pidentify})$ reads
\begin{eqnarray}
\nonumber
\pui{sb}{\psi_{1}}{\psi_{2}}&=&
\eta_{1}\bra{\Psi_{1}}\E^{sb}_{1}\ket{\Psi_{1}}_{ABC}
+ \eta_{2}\bra{\Psi_{2}}\E^{sb}_{2}\ket{\Psi_{2}}_{ABC}
\\ &=&
\frac{\eta_{1}c_{1}+\eta_{2}c_{2}}{2}(1-|\Bra{\psi_{1}}\ket{\psi_{2}}|^{2})\, .
\end{eqnarray}
Depending on the prior probabilities $\eta_{1},\eta_{2}$
the values of $c_{1},c_{2}$ maximizing $\pui{sb}{\psi_{1}}{\psi_{2}}$
read:
$c_{1}=0$, $c_{2}=1$ for $\eta_{1}<\eta_{2}$,
$c_{1}=1$, $c_{2}=0$ for $\eta_{1}>\eta_{2}$ and
$c_{1}+c_{2}=1$ for $\eta_{1}=\eta_{2}$. Hence for equal
prior probabilities the swap-based UI measurement is independent of the
particular choice of $c_{1}$ and $c_{2}$, i.e.
\begin{equation}
\label{Prob_identification}
\pui{sb}{\psi_{1}}{\psi_{2}}
=\frac{1}{4}(1-|\Bra{\psi_{1}}\ket{\psi_{2}}|^{2}).
\end{equation}
However due to symmetry reasons we will further consider $c_1=c_2=1/2$
in the case $\eta_{1}=\eta_{2}$, which gives
\begin{equation}
\label{ui_swap_based}
\E^{sb}_{1}= \frac{1}{2} \one_{B}\otimes \as{AC},
\quad \E^{sb}_{2}= \frac{1}{2} \one_{C}\otimes \as{AB} \, .
\end{equation}
Averaging over the Bloch sphere and using the identity
$\int\int_{S_d}|\Bra{\psi_1}\ket{\psi_2}|^2 d\psi_1 d\psi_2=1/d$
\cite{zyckowski}
we obtain the average probability for the swap-based UI machines
\begin{equation}
\mpui{sb}{S_d}=\frac{1}{4}\left(\frac{d-1}{d}\right)\, .
\end{equation}
Although the probabilities itself are independent of the dimension
the average value converges to $1/4$ in the limit of $d\to\infty$.
This corresponds to an intuitive expectation
that two randomly chosen unit vectors in $\Hs_d$ are more likely
to be orthogonal for higher values of $d$.

\subsection{Optimal measurement}
Although POVM elements $\E^{sb}_{1},\E^{sb}_{2}$ proportional to projectors onto the antisymmetric parts of the subsystem AC
respectively AB,  intuitively seem to be the best universal UI measurement, it was shown by A. Hayashi {\it et al.} \cite{hayashi}
that this is not the case for $\eta_1=\eta_2=\eta=1/2$.
The maximization of the average probability
$\mpui{opt}{S_d}$ can be done by exploiting symmetry via joint
representations of the unitary group $U\otimes U\otimes U$ ($U\in U(d)$)
and the permutation group $S(3)$ permuting the subsystems $A$,$B$ and $C$
of $\Hs^{\otimes3}_d$. In our case Hayashi's
optimal POVM measurement can be written explicitly in the form \cite{hayashi}
\begin{eqnarray}
\E^{opt}_1&=&e \one_B\otimes\as{AC}\, ;
\nonumber
\\
\E^{opt}_2 & =& e \one_C\otimes\as{AB}\, ,
\end{eqnarray}
where $e=\sum_\lambda e_\lambda \Gamma_\lambda$ is a specific operator.
The parameter $\lambda$ specifies both $U(d)$ and $S(3)$ irreducible
representation, $\Gamma_\lambda$ is the projector onto that invariant
subspace of $\Hs^{\otimes 3}_d$, and $e_\lambda$ are non-negative real numbers.
In our case only two irreducible $U(d)$ representations specified by
Young tableaux
$\lambda=(2,1,0,\ldots,0)$, $\lambda=(1,1,1,\ldots,0)$ are relevant.
The corresponding $e_\lambda $'s are $2/3$ and $1/2$.
Therefore we have
\begin{eqnarray}
e=\frac{2}{3} \Gamma_{(2,1,0,\ldots,0)}+\frac{1}{2} \Gamma_{(1,1,1,\ldots,0)}.
\end{eqnarray}
The projectors $\Gamma_{(2,1,0,\ldots,0)}$ and $\Gamma_{(1,1,1,\ldots,0)}$ project onto the subspaces $(V_S\oplus V_{AS})^\perp$
and $V_{AS}$, respectively, where $V_S$ ($V_{AS}$) is the totally symmetric (antisymmetric) subspace of $\Hs^{\otimes3}_d$.
Operators $\one_C\otimes\as{AB}$, $\one_B\otimes\as{AC}$ do not mix the subspaces $(V_S\oplus V_{AS})^\perp$ and $V_{AS}$ on which the operator $e$
acts only a multiple of the identity, i.e.
$e|_{V_{AS}}=e|_{(V_S\oplus V_{AS})^\perp}=\one$. Therefore $\E^{opt}_1$ (analogously $\E^{opt}_2$) is essentially
$\frac{2}{3}\one_B\otimes\as{AC}$ except for $V_{AS}$, where it is $\frac{1}{2}\one_B\otimes\as{AC}$.
Furthermore the POVM elements $\E^{opt}_1$, $\E^{opt}_2$ acting on the totally antisymmetric subspace $V_{AS}$
do not contribute to $\pui{opt}{\psi_1}{\psi_2}$ and $\mpui{opt}{S_d}$, because input states $\ket{\Psi_j}_{ABC}$ [ see Eq.~(\ref{stavy1})]
are symmetric in a pair of subsystems. Thus for calculation of probabilities of identification we can as well use the operators
$\E_1=\frac{2}{3}\one_B\otimes\as{AC}=\frac{4}{3}\E^{sb}_1$, $\E_2=\frac{2}{3}\one_C\otimes\as{AB}=\frac{4}{3}\E^{sb}_2$. Hence, in the optimal case
\begin{eqnarray}
\pui{opt}{\psi_1}{\psi_2}& = &\frac{1}{3}(1-|\Bra{\psi_1}\ket{\psi_2}|^{2}) \label{puihayashi}\, , \\
\mpui{opt}{S_d}& = & \frac{1}{3}\frac{d-1}{d} \, .
\end{eqnarray}

\section{Unambiguous Identification of coherent states}
\label{uicoherent}
Unlike previous sections, where we have considered an unambiguous identification of quantum states from
a finite-dimensional Hilbert space $\Hs_d$, here we work with a semi-infinite dimensional Hilbert space of a linear harmonic oscillator $\Hs_\infty$,
which models a single mode of an electromagnetic field (EM).
The techniques presented in Sec.~III for qudits work for any dimension $d$. The resulting POVM elements are expressed via constant multiples
of projectors, which in large $d$ limit define projectors on $\Hs_\infty^{\otimes 3}$. Therefore we have formally the same universal UI
measurement also for states from $\Hs_\infty$ and this measurement is
optimal for the case of equal prior probability $\eta_1=\eta_2$.

Our goal in this section is to show that an unambiguous identification of coherent states can be done with much better probability of
identification than the optimal UI for all pure states from $\Hs_\infty$. The  basic intuition for this is that coherent states form a
very small subset $S_{coh}$ of all pure states from $\Hs_\infty$ and thus there could be a better way to identify them. The more reasonable
motivation is based on the following observation. As it was mentioned in Sec. \ref{relcomparison} for a special choice of the parameters
$\eta_1,\eta_2$ the optimal POVM for qubits coincides with the optimal quantum state comparison measurement. Hence if there is a better
quantum state comparison of coherent states, which can be used to design an UI setup for coherent states then this setup could perform better than
the universal UI measurement identifying all states from $\Hs_\infty$. E. Andersson {\it et al.} \cite{andersson} proposed such a quantum state comparison
setup, which is also simply realizable by a beamsplitter and a photodetector. In what follows we will explain how their setup works.
In addition, we will present a proof that it
performs optimal quantum state comparison of coherent states. Then we will show how it  can be used to design an unambiguous identification setup for
coherent states.

\subsection{Quantum comparison of coherent states}
A coherent state $|\alpha\rangle$ is fully specified by a complex amplitude $\alpha$. It is a pure state $\ket{\alpha}\in \Hs_\infty$, which is an eigenstate of
the annihilation operator $a\ket{\alpha}=\alpha\ket{\alpha}$. A beamsplitter is a passive-optics device acting on a pair of EM field modes.
Its action is described by the Hamiltonian $H= i \theta(ab^{\dagger}-a^{\dagger}b)$ generating the
unitary transformation $U=e^{\theta t(ab^{\dagger}-a^{\dagger}b)}$,
where $a,a^{\dagger},b,b^{\dagger}$ are creation and annihilation operators of the two modes.
The operation of the beamsplitter is particularly simple for coherent states and is determined by the interaction
time, i.e. by the transmittivity $T$ and the reflectivity $R$ of the beamsplitter:
\begin{eqnarray}
\ket{\alpha}\otimes\ket{\beta}\mapsto \ket{\sqrt{T}\alpha+\sqrt{R}\beta}\otimes\ket{-\sqrt{R}\alpha+\sqrt{T}\beta}
 \label{beamtransf}
\end{eqnarray}
with $T+R=1$.
In comparison of coherent states we want to unambiguously distinguish between $\ket{\alpha}=\ket{\beta}$ and $\ket{\alpha}\neq\ket{\beta}$.
This is equivalent to distinguishing $\beta-\alpha=0$ and $\beta-\alpha\neq 0$, which can be done by 50/50 beamsplitter ($T=R=1/2$) in the
following way. The state of the second mode
after passing through the beamsplitter will be either the vacuum $\ket{0}$ or the state $\ket{\frac{1}{\sqrt{2}}(\beta-\alpha)}$ when $\alpha\neq\beta$.
Thus if we detect at least one photon in the second mode, which happens with probability
$1-|\Bra{0}\ket{\frac{1}{\sqrt{2}}(\beta-\alpha)}|^2=1-e^{-\frac{1}{2}|\alpha-\beta|^2}$, we are sure that the states were different.
The detection of no photons is inconclusive, because all coherent states have a nonzero overlap with the vacuum.

Now we are going to prove optimality of this setup serving as unambiguous coherent state comparator.
First we derive the optimal quantum state comparison measurement for coherent states. In general it is a POVM with two elements $\C^{opt}_0$
(inconclusive result) and $\C^{opt}_1$ (unambiguously indicating the inequality of states)
obeying the following equations:
\begin{eqnarray}
{\rm Tr} \Big[\C^{opt}_1\ket{\alpha}_X\bra{\alpha}\otimes\ket{\alpha}_Y\bra{\alpha}\Big]=0\, ; \ \ \ \ \forall\alpha  \label{cnerror}\, , \\
\C^{opt}_0\geq 0,\quad \C^{opt}_1\geq 0\, ;\quad \C^{opt}_0+\C^{opt}_1=\one. \label{comppovm}
\end{eqnarray}
Integrating the no-error condition (\ref{cnerror})
over all coherent states we obtain the following condition
\begin{eqnarray}
0&=&\int_\complex d\alpha
{\rm Tr}\Big[\C^{opt}_1\ket{\alpha}\bra{\alpha}
\otimes\ket{\alpha}\bra{\alpha}\Big]
={\rm Tr}[\C^{opt}_1 \Delta]
\label{nerorint}
\end{eqnarray}
which defines the operator $\Delta$.
The no-error conditions given by (\ref{cnerror}) are  completely equivalent to Eq.~(\ref{nerorint}),
because the trace under the integral is nonnegative. The operators $\C^{opt}_1$ and $\Delta$ are positive and therefore
Eq.~(\ref{nerorint}) means that these two operators have orthogonal supports. Hence the largest possible support the operator
$\C^{opt}_1$ can have is the orthogonal complement to the support of $\Delta$. It is possible to show that
\begin{eqnarray}
\label{delta_op}
\Delta&=&\frac{\pi}{2}\sum^{\infty}_{N=0} (\ket{\chi_N}\bra{\chi_N}), \label{deltachi}
\end{eqnarray}
where the vectors $\ket{\chi_N}=2^{-\frac{N}{2}}\sum^{N}_{k=0}
\sqrt{\binom{N}{k}} \ket{k}\otimes \ket{N-k}$
are mutually orthonormal, i.e. $\Bra{\chi_N}\ket{\chi_M}=\delta_{N,M}$
The calculation of $\Delta$ is described in detail in Appendix B.

Moreover the normalized operator $\frac{2}{\pi}\Delta$ is a projector with
the same support as $\Delta$. The support of the projector
$\one-\frac{2}{\pi}\Delta$  is therefore the largest possible support
of $\C^{opt}_1$. The optimal measurement must maximize the probability
of revealing the difference of the states launched into the comparator:
\begin{eqnarray}
P_R(\ket{\alpha},\ket{\beta})={\rm Tr}[\C^{opt}_1 \ket{\alpha}\bra{\alpha}\otimes\ket{\beta}\bra{\beta}]\, ,
\end{eqnarray}
while keeping the positivity ($0\leq \C^{opt}_1 \leq\one$)
and the no-error conditions satisfied. Combining these two conditions
on the support of $\C^{opt}_1$ yields $\C^{opt}_1=\one-\frac{2}{\pi}\Delta$.
Thus the optimal unambiguous coherent state comparison
is acomplished by the following projective measurement
\begin{eqnarray}
\nonumber
\C^{opt}_0&=&\frac{2}{\pi}\Delta = \sum^{\infty}_{N=0} (\ket{\chi_N}\bra{\chi_N})_{XY}\, ;\\
\C^{opt}_1&=&\one-\sum^{\infty}_{N=0} (\ket{\chi_N}\bra{\chi_N})_{XY}\, .
\label{cmopt}
\end{eqnarray}

Now it is sufficient to show that the described quantum state comparison setup performs the measurement (\ref{cmopt}).
The mathematical description of the setup is simple. First the beamsplitter acts on the modes $X$ and $Y$ via a unitary transformation
$U(\theta=\pi/4)$ and then the photodetector discriminate between the zero-registered-photon result $\C^{d}_0=\one_X \otimes(\ket{0}\bra{0})_Y$ and
the at-least-one-registered
photon $\C^{d}_1=\one_X \otimes(\one-\ket{0}\bra{0})_Y$ in the mode Y. Hence the measurement performed by the setup is given as
\begin{eqnarray}
\C^{bs}_j=U^\dagger \C^{d}_j U \, ;\quad j=0,1\, .
\end{eqnarray}
Unitarity of $U$ guarantees that $\C^{bs}_j$ form a projective measurement as well as $\C^{d}_j$. Therefore it suffice to show the equality $\C^{bs}_0=\C^{opt}_0$. It follows that
\begin{eqnarray}
\C^{bs}_0=\sum^\infty_{N=0} U^\dagger (\ket{N}_X\otimes\ket{0}_Y{}_X\bra{N}\otimes{}_Y\bra{0}) U
\end{eqnarray}
and the  proof is concluded by showing that $U^\dagger \ket{N}_X\otimes\ket{0}_Y=\ket{\chi_N}_{XY}$.
The technical details are placed in Appendix C.

\subsection{UI measurement with beamsplitters}
For the case of a qubit (Sec.~II)
we have seen explicitly that the unambiguous
identification is very closely related to the problem of the state comparison.
In some sense the identification seems to consist of two state
comparisons performed somehow simultaneously in a single run. However,
because only a single copy of the unknown state is available,
a very specific ``cloning'' machine should be used in order to make such
reduction of the identification problem possible. Usually the cloning
machines \cite{buzek1996} distributes the original
quantum information (represented by quantum state)
among several quantum systems \cite{rosko}. Such approach results
in a complicated entangled state such
that the individual systems are described by density matrices on average
``closest'' to the original quantum state. Unfortunately such cloning cannot
be used for our purposes. The potential clones
described by mixed quantum states cannot be unambiguously
compared with pure states. Therefore we need a very specific cloning
machine producing the clones in pure states, i.e.
$\ket{\psi}\otimes\ket{0}\otimes\ket{0}_{\rm ancilla}
\to\ket{\psi^\prime}\otimes\ket{\psi^\prime}\otimes
\ket{\phi(\psi)}_{\rm ancilla}$. The cloning of coherent states
has been analyzed in \cite{braunstein}, where it was shown that
the single beamsplitter assisted by a linear amplifier
is optimal. Without the linear amplifier the beamsplitter alone
performs on coherent states the transformation
$\ket{\alpha}\otimes\ket{0}\to\ket{\sqrt{T}\alpha}\otimes\ket{\sqrt{R}\alpha}$,
where $R,T$ stands for reflectivity and transmittivity
of the beamsplitter. And this is exactly a type of cloning we are looking
for. Hence, the idea is to use one beamsplitter to clone the system $A$
into two modes and afterwards use another two beamsplitters for particular
state comparisons. Hence, in addition to the modes A,B,C we add an ancillary
mode D set initially to vacuum, i.e.
$\ket{\Phi_{in}}=\ket{\nstav}_A\otimes\ket{\alpha_1}_B
\otimes \ket{\alpha_2}_C \otimes \ket{0}_D$, where $\ket{\nstav}$
is guaranteed to be either $\ket{\alpha_1}$ or $\ket{\alpha_2}$.

Our setup is composed of three beamsplitters the action of which
is described by the unitary transformation
\begin{eqnarray}
\ket{\Phi_{in}}\mapsto U_{3}(DC) U_{2}(BA) U_{1}(DA)\ket{\nstav}\ket{\alpha_1}\ket{\alpha_2}\ket{0}\, ,
\end{eqnarray}
where $U_{j}(XY)$ is associated with the $j$-th
beamsplitter acting on the modes X and Y.
The first beamsplitter $B_1$ (with the transmittivity $T_1$) prepares two clones of the
unknown state $|\alpha_j\rangle_A$ that is encoded in the mode $A$
\begin{eqnarray}
\ket{0}_D\otimes\ket{\nstav}_A \mapsto \ket{\sqrt{R_1}\nstav}_D\otimes\ket{\sqrt{T_1}\nstav}_A.
\end{eqnarray}
The output system remains in a product state, hence
the beamsplitters $B_2$ and $B_3$ can be analyzed separately
\begin{eqnarray}
\label{bs2stav}
B_2:\ket{\alpha_1}_B\otimes\ket{\sqrt{T_1}\nstav}_A &\mapsto&
\ket{\sqrt{T_2}\alpha_1+\sqrt{R_2 T_1}\nstav}_B\\
& \otimes & \ket{-\sqrt{R_2}\alpha_1+\sqrt{T_2 T_1}\nstav}_A \, ,
\nonumber
\end{eqnarray}
and
\begin{eqnarray}
\label{bs3stav}
B_3:\ket{\sqrt{R_1}\nstav}_D\otimes\ket{\alpha_2}_C&\mapsto&
\ket{\sqrt{T_3 R_1}\nstav+\sqrt{R_3}\alpha_2}_D\\
\nonumber
& \otimes &\ket{-\sqrt{R_3 R_1}\nstav+\sqrt{T_3}\alpha_2}_C.
\end{eqnarray}
In case $\nstav=\alpha_1$ we want that the beamsplitters $B_2,B_3$
behave as in the comparison protocol of identical states
$\ket{\alpha_1}$, $\ket{\alpha_1}$, i.e. the modes $A,C$, respectively,
should be transformed into vacuum. Such conditions tell us how the
parameters of the beamsplitters should be adjusted, in particular, we obtain
identities
\begin{eqnarray}
T_2=\frac{1}{1+T_1}\, ; \quad T_3=\frac{1-T_1}{2-T_1}\, ,
\label{t2set}
\end{eqnarray}
where we used the identity $T_j+R_j=1$.

The conditions specified by Eqs.(\ref{t2set}) can be met simultaneously, therefore
we set the transmittivities $T_2,T_3$ accordingly.
The final state of our four modes after passing all three
beamsplitters can be simply obtained from Eqs. (\ref{bs2stav}-\ref{bs3stav}) and reads
\begin{eqnarray}
\nonumber
\ket{\Phi_{out}}&= &
\ket{\sqrt{R_2}(\nstav-\alpha_1)}_A
\otimes\ket{\sqrt{T_2}\alpha_1+\sqrt{R_2 T_1}\nstav}_B
\\ \nonumber
& \otimes&\ket{\sqrt{T_3}(\alpha_2-\nstav)}_C
\otimes\ket{\sqrt{T_3 R_1}\nstav+\sqrt{R_3}\alpha_2}_D \, .
\end{eqnarray}

The field modes are still factorized and
we can focused only on states of modes $A$ and $C$ that are
detecting whether the unknown state matches with $\alpha_1$,
or $\alpha_2$. Indeed, depending on $\nstav$ the modes $A$ and $C$
end up in the states
\begin{eqnarray}
\nstav=\alpha_1 : \quad \ket{0}_A
\otimes\ket{\sqrt{T_3}(\alpha_2-\alpha_?)}_C \, ;\nonumber\\
\nstav=\alpha_2 : \quad \ket{\sqrt{R_2}(\alpha_?-\alpha_1)}_A
\otimes\ket{0}_C \, .
\label{identifstavy}
\end{eqnarray}
Measuring photon number in the modes $A$ and $C$
by photodetectors $P_2$ and $P_1$, respectively, we can
unambiguously identify the unknown state.
In each single run of the experiment we can distinguish four situations:
{\it i)} none of the detectors click,
{\it ii)} only $P_1$ clicks,
{\it iii)} only $P_2$ clicks,
{\it iv)} both detectors click.
In our situation both detectors cannot click
at the same time, because at least one of the modes is in the vacuum.
If only the detector $P_1$ clicks from Eqs. (\ref{identifstavy})
we unambiguously conclude that $\nstav=\alpha_1$.
Similarly if only the detector $P_2$ clicks we unambiguously conclude
that $\nstav=\alpha_2$. If none of the detectors click we cannot
determine which mode was not in the vacuum and therefore it is
an inconclusive result.

In the case $\nstav=\alpha_1$ the probability of a correct identification
follows from equations (\ref{identifstavy}) and is given by the probability
of detecting at least one photon in the mode C
\begin{equation}
P_1=1-|\Bra{0}\ket{\sqrt{T_3}(\alpha_2-\alpha_1)}|^2
=1-e^{-\frac{1-T_1}{2-T_1}|\alpha_1-\alpha_2|^2}\, .
\end{equation}
In case $\nstav=\alpha_2$ the probability of a correct
identification is given by the probability of detecting at least
one photon in the mode A
\begin{equation}
P_2=1-|\Bra{0}\ket{\sqrt{R_2}(\alpha_2-\alpha_1)}|^2
=1-e^{-\frac{T_1}{1+T_1}|\alpha_1-\alpha_2|^2}\, .
\end{equation}
Thus the total probability of an identification for reference states
$\ket{\alpha_1}$ and $\ket{\alpha_2}$ is equal to
\begin{eqnarray}
\pui{bs}{\alpha_1}{\alpha_2}=\eta_1 P_1+\eta_2 P_2 \, .
\label{pibsstavy}
\end{eqnarray}

Next we want to optimize the performance of the setup by properly
choosing the transmittivity $T_1$. The definition of the uniform
distribution on the set of coherent states is problematic,
therefore we first focus on the probability of identification for
a particular choice of reference states
$\ket{\alpha_1}_B$, $\ket{\alpha_2}_C$ expressed by Eq.~(\ref{pibsstavy}).
In fact, this will later help us to draw more general conclusions.
By plotting the $\pui{bs}{\alpha_1}{\alpha_2}$ for various ranges of
$|\alpha_1-\alpha_2|$,$\eta_1 \in [0,1]$  and $T_1 \in [0,1]$ one quickly finds that for the fixed values of $\eta_1$ and $|\alpha_1-\alpha_2|$
the probability $\pui{bs}{\alpha_1}{\alpha_2}$ is maximal for the values  of $T_1$ that depend on $\eta_1$ and $|\alpha_1-\alpha_2|$.
Thus in general for an
arbitrary prior probabilities the optimal transmittivity $T_1$ depends on the reference states to be identified.
However we will show that in a special case of equal prior probabilities there is only one value of the transmittivity $T_1$,
which is optimal for all reference states. This value turns out to be $T_1=1/2$ as one would expect from  symmetry arguments.
In order to show this we calculate $\frac{\partial \pui{bs}{\alpha_1}{\alpha_2}}{\partial T_1}$  from Eq.~(\ref{pibsstavy})
for $\eta_1=\eta_2=1/2$ and
the condition for critical points (vanishing the first derivative) yields
\begin{eqnarray}
1=\frac{(1+T_1)^2}{(2-T_1)^2}e^{-|\alpha_1-\alpha_2|^2(\frac{1-T_1}{2-T_1}-\frac{T_1}{1+T_1})}
\; .
\label{pifd}
\end{eqnarray}
For $0\leq T_1<1/2$ both terms on the right hand side (rhs) of Eq.~(\ref{pifd}) are greater than 1, for $1/2<T_1\leqq 1$ both
terms are less than 1 and for $T_1=1/2$ both terms on the rhs are equal to unity.
Thus $T_1=1/2$ is the only critical point for all reference states and
because of the second derivative being negative it is the global maximum of $\pui{bs}{\alpha_1}{\alpha_2}$ in the interval
$T_1 \in [0,1]$.

\begin{figure}
\begin{center}
\includegraphics[width=7cm]{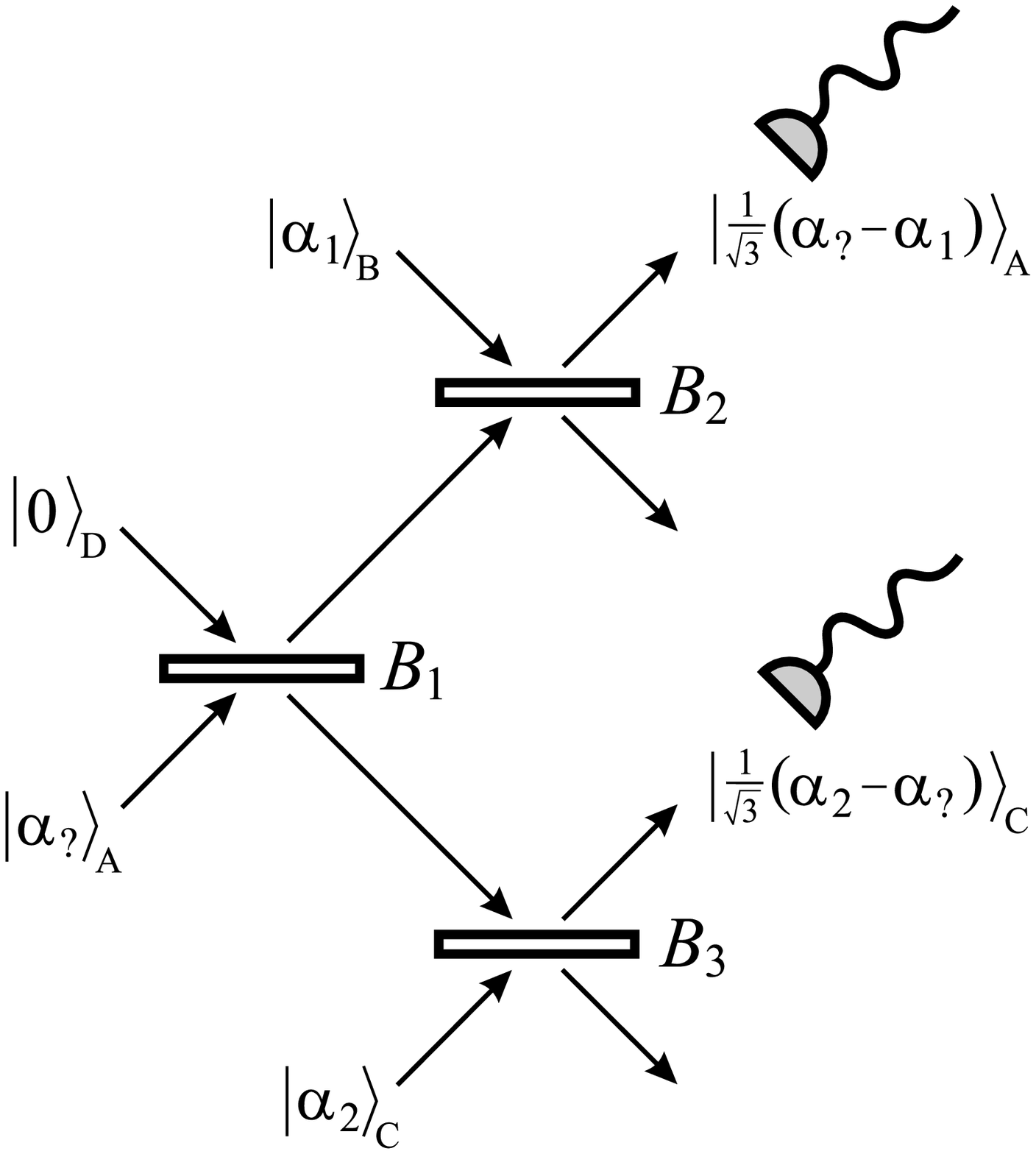}
\caption{The beamsplitter setup designed for an unambiguous identification  of
coherent states.}
\label{nsetup}
\end{center}
\end{figure}

Further, we will consider the UI of coherent states appearing with
equal prior probabilities. In this case the optimal choice of transmittivities
for our three beamsplitter setup is $T_1=1/2$, $T_2=2/3$, $T_3=1/3$.
As a result we obtain that the probability of an unambiguous identification
(\ref{pibsstavy}) reads
\begin{eqnarray}
\pui{bs}{\alpha_1}{\alpha_2}
=1-e^{-\frac{1}{3}|\alpha_1-\alpha_2|^2} \, .
\label{puibs}
\end{eqnarray}

\subsection{Swap-like UI design for coherent  states}
In this section we will propose different unambiguous identification
measurement for coherent states. This approach will be essentially
the same as in the construction of the swap-based UI measurement for qudits, i.e.
also motivated by the state-comparison problem.
The difference is that instead of considering all states we will
be restricted to coherent states only, i.e. the role of an antisymmetric
subspace is played by the projector
$\one-\sum^{\infty}_{N=0}\ket{\chi_N}\bra{\chi_N})_{AC}$, which
was crucial for the optimal state comparison of coherent states (\ref{cmopt})
discussed in Sec.~IV.A. Like before the conclusive POVM
elements $E_j^{sbf}$ ignore the identified mode and conclusively
compare the states in the other two modes, i.e. reading that the states
of these two modes are different. Such a POVM has the following
structure
\begin{eqnarray}
\E^{sbf}_{1}&=& c_{1} \one_{B}\otimes (\one-\sum^{\infty}_{N=0}\ket{\chi_N}\bra{\chi_N})_{AC}\, ; \nonumber \\
\E^{sbf}_{2}&=& c_{2} \one_{C}\otimes (\one-\sum^{\infty}_{N=0}\ket{\chi_N}\bra{\chi_N})_{AB}\, ; \nonumber \\
\E^{sbf}_{0}&=& \one-\E^{sbf}_{1}-\E^{sbf}_{2}\, .
\label{ui_swap_based_fashion}
\end{eqnarray}

Let us fix the parameters $c_1$, $c_2$ and calculate
$\pui{sbf}{\alpha_1}{\alpha_2}$ for this UI measurement.
Using the identity
$\sum^{\infty}_{N=0}\ket{\chi_N}\bra{\chi_N}
=\frac{2}{\pi}\int_\complex \ket{\gamma}\bra{\gamma}\otimes\ket{\gamma}
\bra{\gamma} d\gamma$ [see Eq.~(\ref{deltachi})] we obtain
\begin{eqnarray}
\label{puisbf1}
\pui{sbf}{\alpha_1}{\alpha_2}&=&(\eta_1 c_1 + \eta_2 c_2)\\
\nonumber
& \times&(1-\frac{2}{\pi}\int_\complex |\Bra{\alpha_1}\ket{\gamma}|^2 |\Bra{\alpha_2}\ket{\gamma}|^2 d\gamma)\, .
\end{eqnarray}
Taking into account that $|\Bra{\alpha}\ket{\gamma}|^2=e^{-|\alpha-\gamma|^2}$
and the rectangular identity $|a|^2+|b|^2=\frac{1}{2}(|a+b|^2+|a-b|^2)$
we can express the integral $I=\int_\complex
|\Bra{\alpha_1}\ket{\gamma}|^2 |\Bra{\alpha_2}\ket{\gamma}|^2 d\gamma$
as follows
\begin{eqnarray}
I&=&e^{-\frac{1}{2}|\alpha_1-\alpha_2|^2} \int_\complex e^{-\frac{1}{2}|\alpha_1+\alpha_2-2\gamma|^2} d\gamma  \nonumber\\
&=&e^{-\frac{1}{2}|\alpha_1-\alpha_2|^2} \int_\complex e^{-2|\gamma|^2} d\gamma=\frac{\pi}{2}e^{-\frac{1}{2}|\alpha_1-\alpha_2|^2} \, .
\label{puisbf2}
\end{eqnarray}
Combining Eqs.~(\ref{puisbf1}) and (\ref{puisbf2}) the unambiguous
identification probability reads
\begin{eqnarray}
\pui{sbf}{\alpha_1}{\alpha_2}=(\eta_1 c_1 + \eta_2 c_2)(1-e^{-\frac{1}{2}|\alpha_1-\alpha_2|^2})\, .
\end{eqnarray}

The positivity of the POVM elements $\E^{sbf}_{1}$, $\E^{sbf}_{2}$ is guaranteed by setting $c_1$, $c_2$ to be nonnegative.
The operator $\E^{sbf}_{0}$ has a block diagonal structure in the basis consisting of number states ordered with respect to the increasing global
photon number. Finding the eigenvalues of this matrix is a difficult problem. Nevertheless, the POVM elements $E_1^{sbf}$ and $E_2^{sbf}$
are proportional to mutually overlapping projectors. Let $\ket{\mu}_{ABC}$ be a vector from supports of both these projectors.
For example, $\ket{\mu}$ can be a vector from the totally antisymmetric subspace. Then $\bra{\mu}E_0^{sbf}\ket{\mu}=1-c_1-c_2$.
Thus, the positivity implies that $c_1+c_2\le 1$. For equal prior probabilities $\eta_1=\eta_2=1/2$ we obtain
\begin{eqnarray}
\pui{sbf}{\alpha_1}{\alpha_2}\leq\frac{1}{2}(1-e^{-\frac{1}{2}|\alpha_1-\alpha_2|^2}) \, .
\label{puisbf}
\end{eqnarray}

\subsection{Comparison of UI measurements for coherent states}
In the previous sections we have discussed four different UI measurements that
can be used to identify coherent states:\newline
{\it i)} the swap-based measurement,\newline
{\it ii)} the optimal measurement,\newline
{\it iii)} the swap-like  measurement,\newline
{\it iv)} the beamsplitter setup.

The first two schemes  unambiguously identify arbitrary states of qudits
in arbitrary dimensions. The remaining two are designed to identify only
coherent states. Although the comparison is usually
understood in terms of average probabilities, we will adopt a different
comparison method evaluating the performance directly
in terms of probabilities $P(\ket{\alpha_1},\ket{\alpha_2})$ for all
pairs of states. It turns out that for all the measurements
these probabilities depend only on a scalar product of the states under consideration.

\begin{figure}
\begin{center}
\includegraphics[width=8cm]{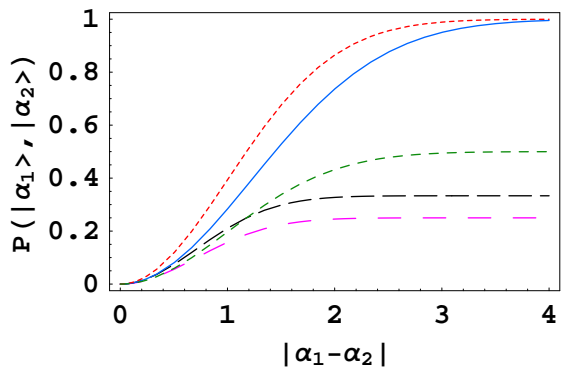}
\caption{(Color online) The probability
of identification $P(|\alpha_1\rangle,|\alpha_2\rangle)$ as
a function of the scalar product (given by $|\alpha_1-\alpha_2|$)
for four UI strategies applied on coherent
states $\ket{\alpha_1}$, $\ket{\alpha_2}$.
Starting from the bottom the two lowest lines  correspond to universal
UI measurements (the swap-based is in magenta (lowest line)
and the optimal strategy is in black, respectively).
The next two lines are associated with the UI measurements designed
for coherent states (the swap-like measurement is in green while the
and the three beamsplitters setup is in solid blue, respectively).
The top (red) curve corresponds to the optimal discrimination
probability among two known states.
}
\label{puigraph}
\end{center}
\end{figure}

As we mentioned at the beginning of this section qudit POVM elements $\E^{sb}_j$ and $\E^{opt}_j$ in the large-$d$ limit
define also POVM elements in $\Hs_\infty$. For simplicity we use the same notation for these operators.
These two UI strategies are universal, so they work for any pure states from $\Hs_\infty$.
If applied on coherent states the corresponding probabilities are given by Eqs. (\ref{Prob_identification}) and (\ref{puihayashi})
\begin{eqnarray}
\nonumber
\pui{sb}{\alpha_1}{\alpha_2}=\frac{1}{4}(1-|\Bra{\alpha_1}\ket{\alpha_2}|^{2})=\frac{1}{4}(1-e^{-|\alpha_1-\alpha_2|^2})\, ; \\
\nonumber
\pui{opt}{\alpha_1}{\alpha_2}=\frac{1}{3}(1-|\Bra{\alpha_1}\ket{\alpha_2}|^{2})=\frac{1}{3}(1-e^{-|\alpha_1-\alpha_2|^2})\, .
\end{eqnarray}
In what follows we will compare
$\pui{sb}{\alpha_1}{\alpha_2}$, $\pui{opt}{\alpha_1}{\alpha_2}$, and
$\pui{bs}{\alpha_1}{\alpha_2}$, $\pui{sbf}{\alpha_1}{\alpha_2}$
which are probabilities of the identification in the UI strategies
designed especially for coherent states [see Eqs. (\ref{puibs}) and
(\ref{puisbf})]. The following inequalities hold for arbitrary
coherent states $\ket{\alpha_1}$ and $\ket{\alpha_2}$
\begin{eqnarray}
P_{sb}\leq P_{sbf}\leq P_{bs} \, ;\quad  P_{opt} \leq P_{bs}\, .
\label{cohineq}
\end{eqnarray}
Hence the same relations hold between the measurements also on average.
They can all be derived in the same way.
Let us define $x=e^{-|\alpha_1-\alpha_2|^2}$ ($x \in [0,1]$)
and $\partial_x=\frac{\partial}{\partial x}$.
All the probabilities are zero for $\alpha_1=\alpha_2$ ($x=1$)
as they should, because in such case reference states coincide.
The validity of these inequalities can be proved
by showing the reversed inequalities for the first derivatives
of probabilities (\ref{cohineq}) with respect to $x$, i.e.
\begin{eqnarray}
\nonumber
& \partial_x P_{sb} \geq\partial_x P_{sbf} \geq\partial_x P_{bs}\, ;
\quad
\partial_x P_{opt}\geq\partial_x P_{bs}
 &
\\ \nonumber &\Leftrightarrow &\\ \nonumber
& -\frac{1}{4} \geq -\frac{1}{2 x^{\frac{1}{2}}}
\geq -\frac{1}{3 x^{\frac{2}{3}}}\,;  \quad -\frac{1}{3}\geq -\frac{1}{3 x^{\frac{2}{3}}}\, . &
\end{eqnarray}
The second row of inequalities obviously holds in the interval $x \in [0,1]$,
so the inequalities (\ref{cohineq}) are proved. More quantitative insight is
given in Fig.(\ref{puigraph}) showing the dependence of the probability
of identification for the considered UI measurements
on the value of $|\alpha_1-\alpha_2|$.

As a result we can conclude that the beamsplitter setup
designed for an unambiguous identification of coherent states
performs better than  other devices including the optimal
universal UI measurement. Another remarkable feature is
that the beamsplitter setup attains $\pui{bs}{\alpha_1}{\alpha_2}=1$
for large values of $|\alpha_1-\alpha_2|$, i.e. in the limit
when two coherent states are orthogonal.

\subsection{Unambiguous identification of $N$ reference states: A quantum database}
Now let us consider  the problem of unambiguously identification among $N$
coherent states $\ket{\alpha_1},\dots,\ket{\alpha_N}$. Our aim is
to modify the proposed beamsplitter scheme to address this slightly
more general problem. In accordance with the case $N=2$ we will
firstly use a beamsplitter array to distribute (to ``copy'') the unknown state
$\ket{\alpha_?}$ onto $N-1$ ancillary modes (which are initially set to the vacuum).
After this redistribution of the information we will {\em simultaneously} implement $N$-fold state comparison
to unambiguously identify the unknown state.

\begin{figure}
\begin{center}
\includegraphics[width=9cm]{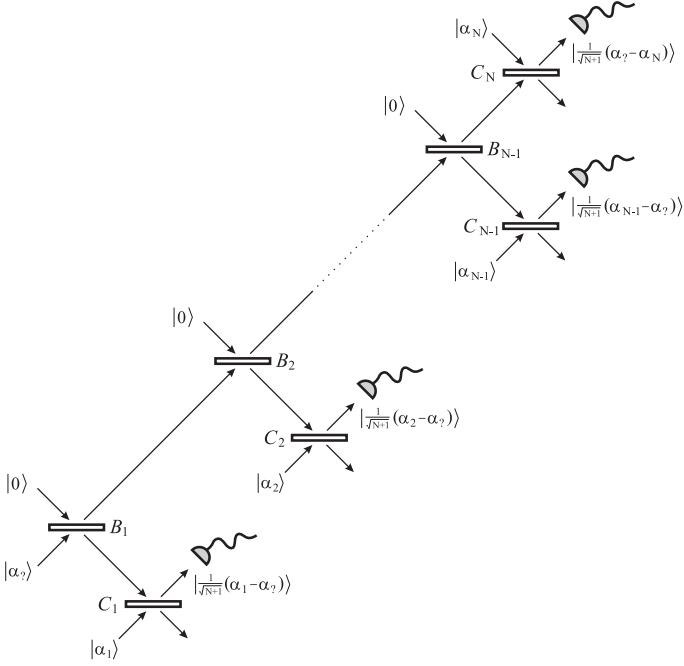}
\caption{Unambiguous identification measurement setup identifying among
$N$ coherent states.}
\label{delic}
\end{center}
\end{figure}

The quantum state distribution can be done with $N-1$ beamsplitters
$B_j$ (described by parameters $R_j,T_j$)
acting on the $j$-th ancillary mode and the mode of an unknown state.
The beasmplitters are applied sequentially splitting the unknown state
into $N$ modes (see Fig.~\ref{delic}) so that each of them end up
in the state $\ket{\frac{1}{\sqrt{N}}\nstav}$. After a little algebra
one can derive the following values for reflectivities and transmittivities
of the $j$-th beamsplitter
\begin{eqnarray}
\nonumber
T_j&=&\frac{N-j}{N-j+1}\, ;\\
R_j&=&\frac{1}{N-j+1}\, .
\end{eqnarray}
Altogether these beamsplitters will implement the transformation
\begin{eqnarray}
\ket{0}^{\otimes N-1}\otimes\ket{\nstav}
\mapsto \ket{\frac{1}{\sqrt{N}}\nstav}^{\otimes N}\, .
\end{eqnarray}

After this transformation is completed we will simultaneously apply
$N$ beamsplitters performing the quantum state comparison of
states $\ket{\frac{1}{\sqrt{N}} \alpha_?}$ and $\ket{\alpha_j}$.
Let us denote by $C_j$ the beamsplitter comparing
the unknown state with the state of the $j$-th mode. Each of them
performs the following transformation
\begin{equation}
\nonumber
\begin{array}{rcl}
\ket{\frac{1}{\sqrt{N}}\nstav,\alpha_k}
&\to &
\ket{\sqrt{\frac{T^c_k}{N}}\nstav+\sqrt{R^c_k}\alpha_k,
-\sqrt{\frac{R^c_k}{N}}\nstav+\sqrt{T^c_k}\alpha_k}\\
&\to &
\ket{\sqrt{\frac{1}{N(N+1)}}(\nstav+N\alpha_k),
\frac{1}{\sqrt{N+1}}(\alpha_k-\nstav)}\, ,
\end{array}
\end{equation}
where we used $R^c_k=\frac{N}{N+1},T^c_k=\frac{1}{N+1}$
and the notation $\ket{\alpha,\beta}$ for $\ket{\alpha}\otimes\ket{\beta}$.
As before the photodetectors monitor the photon number
only in the modes originally in states $\ket{\alpha_j}$ that at the output are
in the state
\begin{equation}
[\otimes^{N-1}_{j=1} \ket{\frac{1}{\sqrt{N+1}}(\alpha_?-\alpha_j)}]\otimes
\ket{\frac{1}{\sqrt{N+1}}(\alpha_N-\alpha_?)}\, .
\end{equation}
Hence if $\ket{\alpha_?}=\ket{\alpha_k}$ the $k$-th mode
is in the vacuum state and all other modes are excited (i.e., populated
by photons). Therefore if all photodetectors except a single one click
then we can unambiguously conclude that the unknown state matches with
the initial state of the mode corresponding to the detector which
did not fire. For all other combination of outcomes the result
is inconclusive. This implies that the probability of unambiguous identification
for the reference states $\ket{\alpha_1},\ldots, \ket{\alpha_N}$ is equal to
\begin{eqnarray}
P(\ket{\alpha_1},\ldots, \ket{\alpha_N})=\sum^{N}_{j=1}\eta_j\prod^{N}_{k\neq j} (1-e^{-\frac{1}{\sqrt{N-1}}|\alpha_k-\alpha_j|^2})\, .
\end{eqnarray}

If we set $\eta_j=\frac{1}{N}$ (for all $j$) and
$\alpha_k=\alpha e^{\frac{2\pi i}{N}k}$ then the probability
can be simplified to
\begin{eqnarray}
P(\ket{\alpha_1},\ldots, \ket{\alpha_N})
=\prod^{N-1}_{k=1}
\left[1-e^{-\frac{2\alpha^2}{\sqrt{N-1}}(1-\cos(\frac{2\pi}{N}k))}\right]\, .
\end{eqnarray}

Concluding this section we note that the unambiguous discrimination among $N$ unknown states described above can be considered
as a search in a quantum database composed of $N$ elements, i.e. $N$ different though unknown coherent states $|\alpha_j\rangle$
that are encoded into $N$ modes of an electromagnetic field. We point out that we have only a single copy of each of the states $|\alpha_j\rangle$
so one can not acquire a complete classical knowledge about the state. This set of $N$ states corresponds to a quantum database.
In addition we have the $(N+1)$-st mode of the light field in the state $|\alpha_?\rangle$. The search of the database corresponds to the
task of matching of two modes such that $|\alpha_?\rangle=|\alpha_j\rangle$. So we can say that the two modes are in the same state without knowing
what the state actually is.

\section{Conclusion}
In this paper we have addressed the problem of an unambiguous identification
of unknown coherent states. We have explicitly designed two UI
measurements taking into account the a priori knowledge about a  particular
family of states and compare these measurements with the universal unambiguous identification,
i.e. the UI measurements (either the swap-based or the optimal one)
that can be applied for all states. Our main goal was to design a
simple experimental setup consisting of three beamsplitters (see Fig.1)
that performs best. Finally, we have generalized the problem and proposed the measurement
unambiguously identifying among $N$ coherent state.

The beamsplitter setup was motivated by an intuitive
reduction of the unambiguous identification problem
into specific ``cloning'' task and an unambiguous state comparison.
We have proved that the state comparison originally proposed
in \cite{andersson} is indeed the optimal one (this was implicitly
conjectured in Ref.~\cite{andersson} but was not proved).
It is interesting to compare UI measurements described
in this paper with an UI measurement given as a mixture of two optimal
unambiguous comparison measurements, i.e.
$\overline{E}_1=q I_{B}\otimes F^{\rm dif}_{AC}$,
$\overline{E}_2=q I_{C}\otimes F^{\rm dif}_{AB}$,
$\overline{E}_0=I-\overline{E}_1-\overline{E}_2$.
Let us consider the UI problem for pure states belonging to a set $S$ such that
$\int_S |\psi\rangle\langle\psi|\otimes|\psi\rangle\langle\psi| d\psi
=\Pi(S)$ is a projector. Under such assumption the conclusive
result for an optimal unambiguous comparison of states from $S$
is associated with the positive operator $F^{\rm dif}=I-\Pi(S)$.
If $S=S_d$ is the set of all pure states of $d$-dimensional system
then $F^{\rm dif}=I-\Pi(S_d)=I-\Pi_{\rm sym}={\rm ASym}$. Hence, the
corresponding POVM consists of operators
$\overline{E}_1=\frac{1}{2}I_B\otimes{\rm ASym}_{AC}$,
$\overline{E}_2=\frac{1}{2}I_C\otimes{\rm ASym}_{AB}$, i.e.
(for $d>2$) it is exactly the SWAP-based UI measurement as specified in
Eq.(\ref{ui_swap_based}). Similarly for coherent states we have
$F^{\rm dif}=I-\Pi(S_{\rm coh})=I-\frac{2}{\pi}\Delta$
(see Eq.(\ref{delta_op})) and the mixing of optimal unambiguous
comparison strategies results in the same POVM
as in Eq.(\ref{ui_swap_based_fashion}). Based on our results
we can say that although for purposes of the state comparison problem the
operator $\one-\Pi({\cal S})$ is the optimal solution,
the resulting UI measurement based on mixing of such
optimal unambiguous comparison strategies is not the optimal one.


The proposed beamsplitters setup (see Fig.1) for unambiguous
state identification can be compared with the
measurement proposed in Ref.~\cite{banaszek} discriminating
optimally among two known coherent states. Both of them consists
of three beamsplitters, but arranged differently. An interesting
observation is that the differences between the
probabilities are not very large (see Fig.2)
and even more surprising is the fact that two unknown nearly
orthogonal coherent states can be identified almost perfectly.
For the universal optimal UI measurement (see Fig.2)
there is a significant gap between the probabilities for state
discrimination and state identification.

The proposal of unambiguous identification of coherent states
is extended for an arbitrary number $N$ of reference states.
We proposed an experimental setup consisting of $2N-1$ beamsplitters
for unambiguous identification among $N$ {\em unknown} coherent states.
This setup can be considered as a search in a quantum database.
The elements of the database are {\em unknown} coherent states encoded
in different modes of an electromagnetic field. The task is to specify
the two modes that are excited in the same, though unknown, coherent state.
The analysis of this aspect of unambiguous identification problem
is beyond the scope of this paper and details will be presented elsewhere.

\section*{ACKNOWLEDGMENTS}
This work was supported by the European Union projects QAP,
CONQUEST, by the Slovak Academy of Sciences via the project CE-PI,
and by the projects APVT-99-012304, VEGA and GA\v CR GA201/01/0413.
\appendix
\section{Eigenvalues of $E_0^{sb}$}
The operator $\E_{0}^{sb}$ is defined in Eq.(\ref{sbpovm}). As we have already mentioned in Section IIIA this operator is block diagonal
and consists of
three types of blocks.

\noindent
1. Trivial $\bra{iii}\E_{0}^{sb}\ket{iii}=1$.

\noindent
2. $3 \times 3$ matrix $\bra{\sigma_{1}(iij)}\E_{0}^{sb}\ket{\sigma_{2}(iij)}$:
\begin{multline}
Q_{3}=
\begin{pmatrix}
    &    iij              &     iji           &    jii      \\
iij & 1 - c_{1}/2         &     0             &    c_{1}/2  \\
iji &     0               & 1 - c_{2}/2       &    c_{2}/2  \\
jii &     c_{1}/2         &     c_{2}/2       &    1 - c_{1}/2-c_{2}/2
\end{pmatrix}
\end{multline}
with eigenvalues
\begin{eqnarray}
\lambda^{(3)}_{1}& =& 1\, ; \\
\lambda^{(3)}_{2,3}& =& \frac{2-c_{1}-c_{2}\pm\sqrt{c_{1}^{2}-c_{1}c_{2}+c_{2}^{2}}}{2}
\nonumber
\end{eqnarray}

\noindent
3. $6 \times 6$ matrix
$\bra{\sigma_{1}(ijk)}\E_{0}^{sb}\ket{\sigma_{2}(ijk)}$:
\begin{multline}
Q_{6}=
\begin{pmatrix}
    & ijk & kji & jki & ikj & kij & jik \\
ijk &  X    & c_1/2 &  0    &  0    &  0    & c_2/2 \\
kji & c_1/2 &  X    & c_2/2 &  0    &  0    &  0    \\
jki &  0    & c_2/2 &  X    & c_1/2 &  0    &  0    \\
ikj &  0    &  0    & c_1/2 &  X    & c_2/2 &  0    \\
kij &  0    &  0    &  0    & c_2/2 &  X    & c_1/2 \\
jik & c_2/2 &  0    &  0    &  0    & c_1/2 &  X
\end{pmatrix}\, ,
\end{multline}
where $X=1-\frac{c_{1}}{2}-\frac{c_{2}}{2}$. The corresponding eigenvalues
read
\begin{equation}
\begin{array}{l}
\lambda^{(6)}_{1}=1\\
\lambda^{(6)}_{2}=1-c_{1}-c_{2}\\
\lambda^{(6)}_{3,4}
=\frac{2-c_{1}-c_{2}+\sqrt{c_{1}^{2}-c_{1}c_{2}+c_{2}^{2}}}{2}\\
\lambda^{(6)}_{5,6}
=\frac{2-c_{1}-c_{2}-\sqrt{c_{1}^{2}-c_{1}c_{2}+c_{2}^{2}}}{2} \, .
\end{array}
\end{equation}

For qubits we have $2^{3}$-dimensional Hilbert space and $\E_{0}^{sb}$ is represented by
$8\times 8$ matrix with two $3\times 3$ blocks and two $1\times 1$ blocks. So
for qubits the sufficient condition for positivity of $\E_{0}^{sb}$ reads
$$
2-c_{1}-c_{2}\pm\sqrt{c_{1}^{2}-c_{1}c_{2}+c_{2}^{2}}\geq 0
$$

For qudits, $d>2$, at least one $6\times 6$ block appears in the matrix
of $\E_{0}^{sb}$. The eigenvalues of the $6\times 6$ block satisfy the
following inequality:
$$
\lambda^{(6)}_{1}\geq\lambda^{(6)}_{3,4}\geq\lambda^{(6)}_{5,6}\geq\lambda^{(6)}_{2},
$$
so the sufficient condition for positivity of $\E_{0}^{sb}$ when $d>2$ is
$\lambda^{(6)}_{2}=1-c_{1}-c_{2}\geq 0$.

\section{Calculation of $\Delta$}
In this appendix we will explicitly evaluate the operator
\begin{eqnarray}
\Delta&=&\int_\complex \ket{\alpha}_X\otimes\ket{\alpha}_Y {}_X\bra{\alpha}\otimes {}_Y\bra{\alpha} \quad d\alpha \, .
\end{eqnarray}
First expand the coherent state $\ket{\alpha}$ in the number basis
and use the polar parametrization of the complex plane
\begin{eqnarray}
\ket{\alpha}=e^{-\frac{|\alpha|^2}{2}}
\sum^{\infty}_{k=0}\frac{\alpha^k}{\sqrt{k!}}\ket{k},
\quad \alpha=r e^{i\phi},
\quad d\alpha=r dr d\phi
 \nonumber
\end{eqnarray}
in order to write
\begin{eqnarray}
 \Delta&=&\int_\complex e^{-2|\alpha|^2} \sum^{\infty}_{k,l,m,n=0} \frac{\alpha^{k+l} (\alpha^{\ast})^{m+n}}{\sqrt{k!l!m!n!}}
\ket{k}\bra{m}\otimes \ket{l}\bra{n} d\alpha \nonumber\\
\nonumber
&=&\left(\sum^{\infty}_{k,l,m,n=0}\frac{1}{\sqrt{k!l!m!n!}}
\int^{\infty}_{0} e^{-2r^2}r^{k+l+m+n+1}dr \right)\\
\nonumber
& \times &
\int^{2\pi}_0 e^{i\phi(k+l-m-n)} d\phi \ket{kl}\bra{mn}\, .
\end{eqnarray}
The integration over $\phi$ yields $2\pi \delta_{k+l,m+n}$
and the integration over $r$ after a simple substitution and
usage of $\delta_{k+l,m+n}$ leads to $\Gamma(k+l+1)/2^{k+l}$.
Using the identity $\Gamma(k+l+1)=\sqrt{(k+l)!}\sqrt{(m+n)!}$
we obtain
\begin{eqnarray}
\nonumber
\Delta&=&\frac{\pi}{2}\sum^{\infty}_{k,l,m,n=0}\frac{\sqrt{\binom{m+n}{m}} \sqrt{\binom{k+l}{k}}}{2^{\frac{k+l}{2}} 2^{\frac{m+n}{2}}} \delta_{k+l,m+n} \ket{kl}\bra{mn}\, .
\end{eqnarray}
Defining a new summation index $N=m+n=k+l$ and using $\delta_{k+l,m+n}$
we eliminate one sum to obtain
\begin{eqnarray}
\Delta&=&\frac{\pi}{2}\sum^{\infty}_{N=0}\sum^{N}_{k,m=0}\frac{\sqrt{\binom{N}{k}} \sqrt{\binom{N}{m}}}{2^{\frac{N}{2}} 2^{\frac{N}{2}}}  \ket{k}\ket{N-k}\bra{m}\bra{N-m} \nonumber\\
&=&\frac{\pi}{2}\sum^{\infty}_{N=0} (\ket{\chi_N}\bra{\chi_N})_{XY},
\end{eqnarray}
where we defined mutually orthogonal vectors
\begin{eqnarray}
\ket{\chi_N}=2^{-\frac{N}{2}}\sum^{N}_{k=0}\sqrt{\binom{N}{k}} \ket{k}_X \otimes \ket{N-k}_Y\, ,
\end{eqnarray}
i.e. $\Bra{\chi_N}\ket{\chi_M}=\delta_{N,M}$.

\section{Calculation of $U^\dagger \ket{N}_X\otimes\ket{0}_Y$}
The action of the 50/50 beamsplitter unitary transformation $U$ on the
creation operators $a^\dagger$, $b^\dagger$ of modes X and Y is given by
\begin{eqnarray}
U^\dagger a^\dagger U&=&\frac{1}{\sqrt{2}}(a^\dagger+b^\dagger) \\
U^\dagger b^\dagger U&=&\frac{1}{\sqrt{2}}(-a^\dagger+b^\dagger)
\end{eqnarray}
Let us express the number (Fock) state via the creation operators acting on the vacuum,
i.e. $\ket{N}_X=(a^\dagger)^N / \sqrt{N!} \ket{0}_X$. The vacuum is unaffected
by both $U$ and $U^\dagger$. Having all this in mind we can calculate
the state $\ket{\chi_N}=U^\dagger \ket{N}_X\ket{0}_Y$
\begin{eqnarray}
\nonumber
& & \ket{\chi_N}=
\frac{1}{\sqrt{N!}} (U^\dagger a^\dagger U)^N \ket{0}\ket{0}\\
\nonumber & & =
\frac{1}{2^{\frac{N}{2}}\sqrt{N!}}(a^\dagger +b^\dagger)^N \ket{0}\ket{0}\\
\nonumber & & =
\frac{1}{2^{\frac{N}{2}}}\sum^{N}_{k=0} \sqrt{\frac{k!(N-k)!)}{N!}}
\binom{N}{k} \frac{(a^\dagger)^k}{\sqrt{k!}} \frac{(b^\dagger)^{N-k}}{\sqrt{(N-k)!}} \ket{0}\ket{0}\\
\nonumber & & =
\frac{1}{2^{\frac{N}{2}}}\sum^{N}_{k=0} \sqrt{\binom{N}{k}}
\ket{k}\ket{N-k}.
\end{eqnarray}


\end{document}